\documentclass{aastex631}

\begin{document}
\shorttitle{K. Luke et al}
\shortauthors{K. Luke et al}

\title{Watch the Moon, Learn the Moon: Lunar Geology Research at School Level with Telescope and Open Source Data}

\author {Kevin J. Luke}
\affiliation{Leader, Science Teacher, Science-AI Symbiotic Group, Seven Square Academy, Naigaon, Vasai-Virar, 
Mumbai, India}
\affiliation{Project Student, Tata Institute of Fundamental Research, Mumbai, India}

\author {Abhinav Mishra}
\affiliation{Member, Student, Science-AI Symbiotic Group, Seven Square Academy, Naigaon, Vasai-Virar, 
Mumbai, India}

\author {Vihaan Ghare}
\affiliation{Member, Student, Science-AI Symbiotic Group, Seven Square Academy, Naigaon, Vasai-Virar, 
Mumbai, India}

\author {Shaurya Chanyal}
\affiliation{Member, Student, Science-AI Symbiotic Group, Seven Square Academy, Naigaon, Vasai-Virar, 
Mumbai, India}

\author {Priyamvada Shukla}
\affiliation{Member, Student, Science-AI Symbiotic Group, Seven Square Academy, Naigaon, Vasai-Virar, 
Mumbai, India}

\author {Anushreya Pandey}
\affiliation{Member, Student, Science-AI Symbiotic Group, Seven Square Academy, Naigaon, Vasai-Virar, 
Mumbai, India}

\author {Vaishnavi Rane}
\affiliation{Member, Student, Science-AI Symbiotic Group, Seven Square Academy, Naigaon, Vasai-Virar, 
Mumbai, India}

\author {Ashadieeyah Pathan}
\affiliation{Member, Student, Science-AI Symbiotic Group, Seven Square Academy, Naigaon, Vasai-Virar, 
Mumbai, India}

\author {Parv Vaja}
\affiliation{Member, Student, Science-AI Symbiotic Group, Seven Square Academy, Naigaon, Vasai-Virar, 
Mumbai, India}

\author {Sai Gogate}
\affiliation{Member, Student, Science-AI Symbiotic Group, Seven Square Academy, Naigaon, Vasai-Virar, 
Mumbai, India}

\author {Shreyansh Tiwari}
\affiliation{Member, Student, Science-AI Symbiotic Group, Seven Square Academy, Naigaon, Vasai-Virar, 
Mumbai, India}

\author {Jagruti Singh}
\affiliation{Member, Student, Science-AI Symbiotic Group, Seven Square Academy, Naigaon, Vasai-Virar, 
Mumbai, India}

\author {Dhruv Davda}
\affiliation{Member, Student, Science-AI Symbiotic Group, Seven Square Academy, Naigaon, Vasai-Virar, 
Mumbai, India}

\begin{abstract}
Science-AI Symbiotic Group at Seven Square Academy, Naigaon was formed in 2023 with the purpose of bringing school students to the forefronts of science research by involving them in hands on research. In October 2023 a new project was started with the goal of studying the lunar surface by real-time observations and open source data. Twelve students/members from grades 8, 9, 10 participated in this research attempt wherein each student filled an observation metric by observing the Moon on various days with a Bresser Messier 150mm/1200mm reflector Newtonian telescope. After the observations were done, the members were assigned various zones on the lunar near side for analysis of geological features. Then a data analysis metric was filled by each of students with the help of Lunar Reconnaissance Orbiter Camera's/ LROC's quickmap open access data hosted by Arizona State University. \textbf{In this short paper a brief overview of this project is given.} One example each of observation metric and data analysis metric is presented. This kind of project has high impact for school science education with minimal costs. This project can also serve as an interesting science outreach program for organisations looking forward to popularise planetary sciences research at school level.

\end{abstract}


\section{Introduction} 
The Moon with its mesmerizing presence in the night sky, has captivated human curiosity for centuries. Its phases, surface features, and subtle changes provide a canvas for scientific exploration and observation. In this study, we delve into the realms of lunar inquiry through a student-driven project that combines hands-on observation and data analysis. By merging the traditional method of direct observation with the wealth of information available from NASA's Lunar Reconnaissance Orbiter Camera's (LROC) quickmap data \citep{lroc_quickmap}, our aim is to unravel the mysteries of Earth's celestial companion. 

As the students engaged in the observational phase, meticulously recording their findings in the observational form, they embarked on a journey of discovery. Armed with curiosity and basic tools, they observed the Moon's prominent features such as craters, mare, and mountains. This experiential learning not only fostered a deeper connection with the night sky but also laid the foundation for their subsequent foray into the digital realm of lunar exploration.

There is a vast amount of open source data lying around on internet that can be used effectively in science education at secondary and high school levels. This open source data can be used to study various astrophysical phenomena effectively, hence giving students practical rather than theoretical understanding of the domain of astrophysics and space sciences.

LROC's quickmap \citep{lroc_quickmap} serves as a powerful tool, offering an unprecedented level of detail about the Moon's surface. From high resolution imagery to topographic maps, this resource provides a virtual gateway to the lunar landscape. By guiding students to individually analyze this data we empower them to navigate the wealth of information available, fostering critical thinking and data interpretation skills.

This paper details the methodology employed in our project, highlighting the synergy between traditional observation and modern technology. Through this integrative approach, we not only deepen our understanding of lunar phenomena but also cultivate a holistic and enriching educational experience for the students involved.

In the following sections, we delve into the introduction to the various lunar missions, project's methodology, present example results of the observation and data analysis metrics, and discuss the implications of our project. This study not only contributes to the growing body of knowledge about the Moon but also underscores the value of merging observational traditions with cutting-edge technological tools in science education.

\subsection{About the LRO mission}
\begin{figure}[hb!]
\centering
\includegraphics[width=0.3\textwidth]{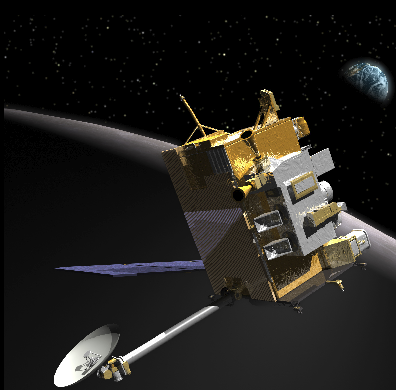}
\caption{An artistic illustration of LRO. Courtesy NASA \citep{lro_nasa}.}

\end{figure}
Lunar Reconnaissance Orbiter (LRO) launched in 2009 by NASA was primarily designed to study the surface of the Moon to enable researchers to create a three dimensional map of the Moon \citep{lro_nasa}. Creating a three dimensional map has major impact on future space exploration missions to the lunar surface. This is due the fact that knowledge of the surface in advance will help space explorers plan the mission with safety in mind in a very efficient way. Additionally the LRO was also designed and armed with sensors to detect the presence of radiation on lunar surface \citep{lro_nasa}. Advance knowledge of these activities will help in timely response in future missions, especially the upcoming Artemis 2 and 3 manned missions. 

At around 4,000 pounds the LRO  was launched in tandem with Lunar Crater Observation and Sensing Satellite (LCROSS) into the lunar orbit by ATLAS V 401 from Cape launch station \citep{lro_nasa}. The LRO hosts a sleuth of instruments for conducting various geological surveys like Cosmic Ray Telescope for the Effects of Radiation (CRaTER), Diviner Lunar Radiometer Experiment (DLRE), Lyman-Alpha Mapping Project (LAMP), Lunar Exploration Neutron Detector (LEND), Lunar Orbiter Laser Altimeter (LOLA), Lunar Reconnaissance Orbiter Camera (LROC) and Mini-RF Miniature Radio Frequency Radar \citep{lro_nasa}.

\subsection{LROC system}

\begin{figure}[h!]
\centering
\includegraphics[width=0.26\textwidth]{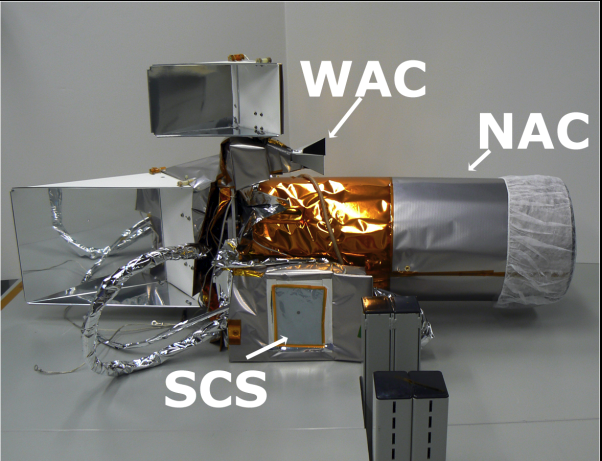}
\hspace{0.02\textwidth}
\includegraphics[width=0.3\textwidth]{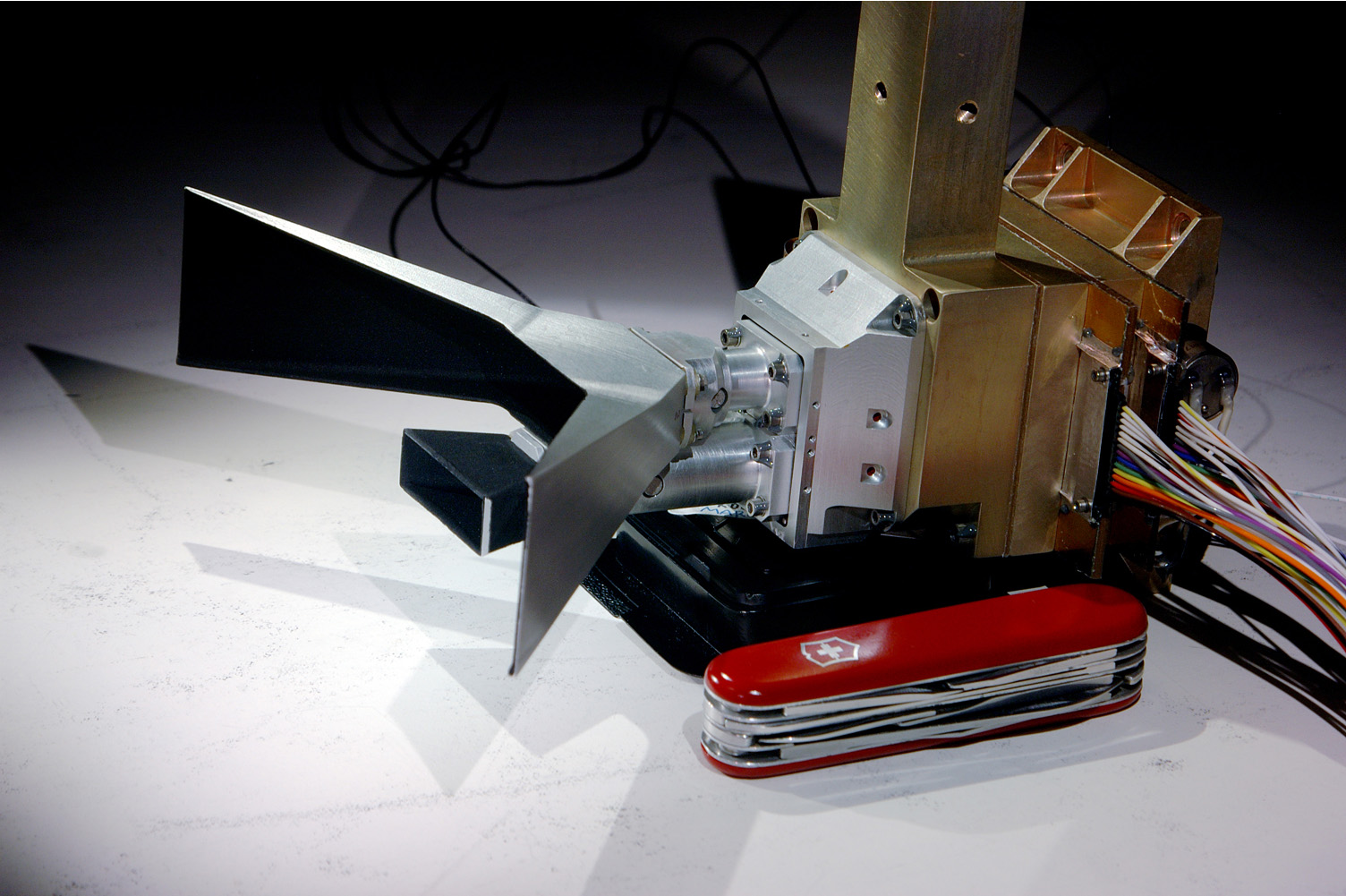}
\hspace{0.02\textwidth}
\includegraphics[width=0.3\textwidth]{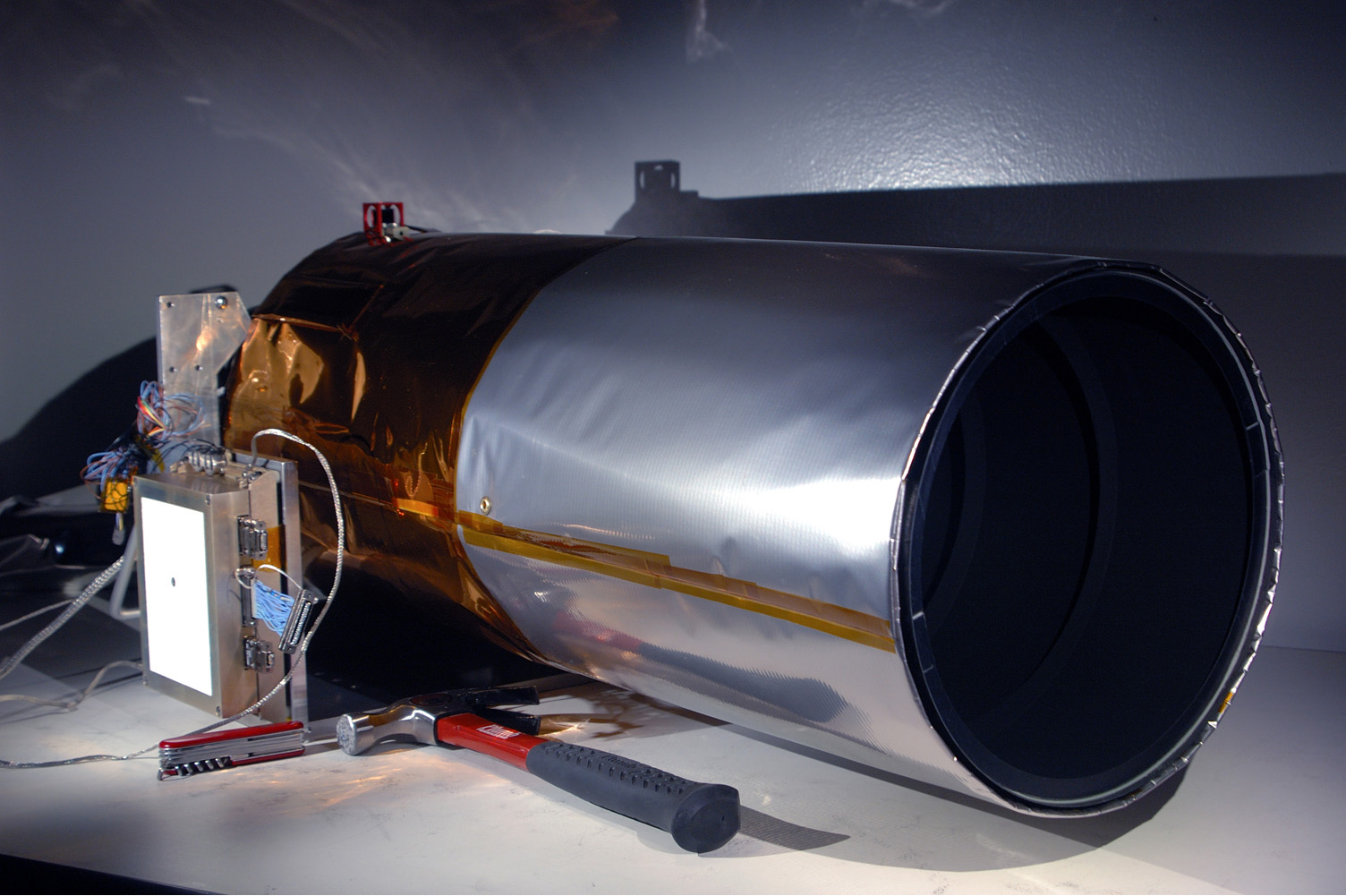}
\caption{The various components of the LROC system. Credits: LROC website \citep{lroc_website_closeup}, Malin Space Science Systems \citep{msss_lro_camera}.}
\end{figure}

The Lunar Reconnaissance Orbiter Camera (LROC) is the most pivotal component onboard LRO with the main objective of recording highly resolved images and to construct a detailed map of the lunar surface \citep{robinson2010lro}. Since its launch on LRO, LROC has been extremely important in advancing lunar exploration by providing extensive high resolution data for scientific analysis and future missions.
As detailed in the Space Review paper \citep{robinson2010lro}, LROC package comprises of the Wide Angle Camera (WAC) for global, wide angled photography of the lunar surface and two Narrow Angle Cameras (NACs) for high resolution photography of the lunar surface with resolutions going down to resolve even the previous manned landing missions. These accurate lunar global and high resolution images stand on a crucially challenging role in identifying safer and science rich landing sites for both manned and autonomous missions, including NASA's Artemis program, which has intended goal of landing humans again on the Moon.

\subsubsection{LROC WAC}

The WAC provides a broader, wide angle view, delivering images at a resolution of 100 meters per pixel across a wide region measuring 105 kilometers in width \citep{msss_lro_camera, robinson2010lro}. The wide angle system employs seven distinct wavelengths (310 - 680 nm) to survey the lunar surface comprehensively. This multi-spectral approach aids in characterizing the distribution of lunar resources, specifically focusing on mineral such as ilmenite, which contains iron, titanium, and oxygen \citep{msss_lro_camera, robinson2010lro}. The detailed spectral analysis enhances our understanding of the Moon's composition and the potential availability of valuable resources. 

\subsubsection{LROC NAC}

The high resolution images are captured with an impressive spatial resolution of 0.5 meters per pixel, offering an unprecedented level of detail with the help of 2 NACs \citep{msss_lro_camera, robinson2010lro}. The NACs creates a swath that is 5 kilometers (3.1 mi) wide, allowing for comprehensive and close-up examinations of specific regions of interest on the Moon. This capability is crucial for identifying surface features, evaluating terrain and aiding in the selection of potential landing sites for future missions.

\subsubsection{LRO Diviner}

The Diviner, a critical component aboard LRO since its launch in 2009, serves the essential function of meticulously mapping the thermal properties of the Moon's surface. Through the precise measurement of infrared radiation emitted by the lunar surface as well as the solar radiation, Diviner creates detailed temperature maps, offering invaluable insights into the Moon's thermal environment through rock abundance and regolith temperature measurements \citep{paige2010lro}. Utilizing an advanced radiometer, Diviner captures thermal emission data across a wide spectral range \citep{paige2010lro}. This data enables the identification of regions with distinct thermal characteristics \citep{paige2010lro, powell2023high}, including the potential presence of water ice in permanently shadowed craters \citep{paige2010lro, powell2023high}. The comprehensive thermal mapping provided by Diviner not only enhances our scientific understanding of the Moon but also plays a crucial role in selecting optimal landing sites and planning resource utilization for future lunar exploration missions. 

For the project, the regolith temperature data is obtained from \citep{powell2023high} through the LROC quickmap \citep{lroc_quickmap}.

\subsection{LRO Laser Altimeter (LOLA)}
Laser altimetry depends on the bombardment of laser pulses on a surface or terrain to enable three dimensional mapping of the surface \citep{barker2015lunar}. By accurate measurements of the return pulses of laser the terrain can be mapped which would lead to the understanding of the elevation and slope characteristics of the surface or terrain. Various spacecrafts have conducted laser altimetry over the years among them being the Clementine mission around the Moon \citep{barker2015lunar}. Operating at a laser frequency of 28 Hz the LOLA instrument maps the terrain elevation of the lunar surface. When the data was combined with the Terrain Camera of Kaguya mission,  high quality terrain elevation maps were produced for the entire lunar surface resulting in global lunar digital elevation model \citep{barker2015lunar}.

The terrain elevation and slope data were obtained from \citep{barker2015lunar, henriksen2020lroc} through the LROC quickmap \citep{lroc_quickmap}.

\subsection{About SELENE/Kaguya Mission}

\begin{figure}[ht!]
\centering
\rotatebox{0}{\includegraphics[width=0.3\textwidth]{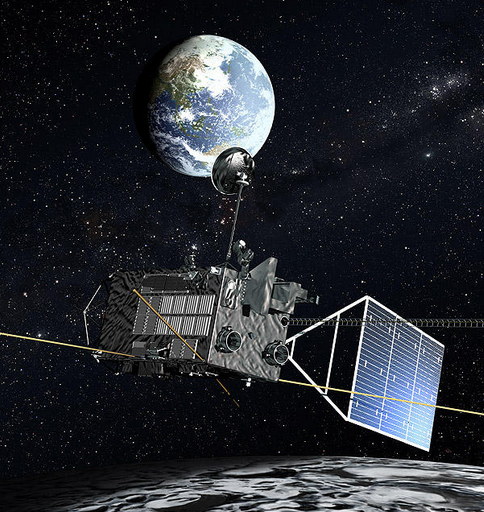}}
\caption{An artistic illustration of SELENE spacecraft. Courtesy:JAXA/Kaguya website \citep{jaxa_kaguya}.}

\end{figure}

The SELENE (Selenological and Engineering Explorer) spacecraft, also known as Kaguya, was a Japanese lunar orbiter mission launched in 2007 with the primary objective of comprehensive lunar exploration. Among its instrumental suite, the Spectral Profiler (SP) and Terrain Camera played a crucial role in determining the abundance of iron oxide as well as the creation of digital elevation maps of the entire lunar surface in combination with the LOLA data from LRO \citep{barker2015lunar, kato2009kaguya, lemelin2019compositions}. The SP, a part of SELENE's scientific payload, employed a combination of visible and infrared spectroscopy to analyze the composition of lunar materials \citep{kato2009kaguya, lemelin2019compositions}. This spectroscopic data allowed scientists to assess the presence and distribution of iron oxide, a key mineral indicative of the Moon's geological history. SELENE's measurements of iron oxide abundance provided valuable insights into the lunar surface's mineralogical composition, contributing to our understanding of geological processes and evolution \citep{kato2009kaguya, lemelin2019compositions}. By mapping the spatial distribution of iron oxide, scientists gained information about the Moon's formation and subsequent volcanic activity \citep{kato2009kaguya}. While SELENE's mission concluded in 2009, the wealth of data it collected, including Terrain Camera data and iron oxide measurements, continues to be a valuable resource for lunar scientists and contributes to the broader understanding of planetary evolution. 

For the project the iron oxide abundance data is obtained from Kaguya iron oxide abundance data \citep{lemelin2019compositions}. The terrain elevation and slope data were obtained from \citep{barker2015lunar, henriksen2020lroc} through the LROC quickmap \citep{lroc_quickmap}.

\section{Methodology} 
For the project a team of twelve students/members were given a number of tasks towards the successful understanding of the lunar surface. The methodology followed is mentioned here.

\begin{enumerate}

    \item An \textbf{observation metric} was prepared for recording of observations of the lunar features by using a personal reflector telescope.
    \item Various members of the team on different nights in the months of October and November did the observations. This was done by him/her by setting up the telescope and then he/she recorded the observations in the observation metric.
    \item Additionally snapshots of the lunar surface were taken by each member using smartphones on the telescope at very high powers available.
    \item After the observations were done, members were assigned  particular areas/zones of the lunar surface for the next step of filling a data analysis metric.
    \item For each zone, an extensive \textbf{data analysis metric} was prepared with respect to  parameters like topography, elevation, slope, regolith temperature and iron oxide abundance.
    \item Each member did the data analysis from the LROC quickmap open access data website and filled the data analysis metric. 
    \item Members took part in podcasts wherein they expressed their student level experiences, ambitions and opinions about lunar research and exploration. The podcast will be available on Youtube. 
\end{enumerate}

\section{Individual Observation}

\begin{figure}[ht!]
\centering
\includegraphics[width=0.2\textwidth]{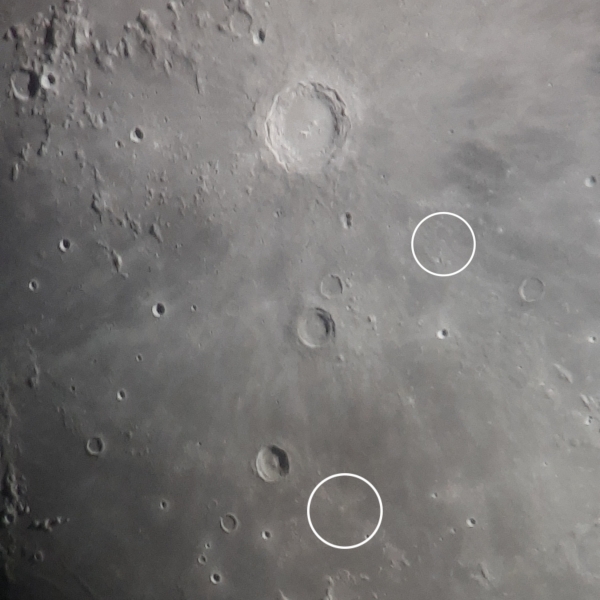}
\hspace{0\textwidth}
\includegraphics[width=0.2\textwidth]{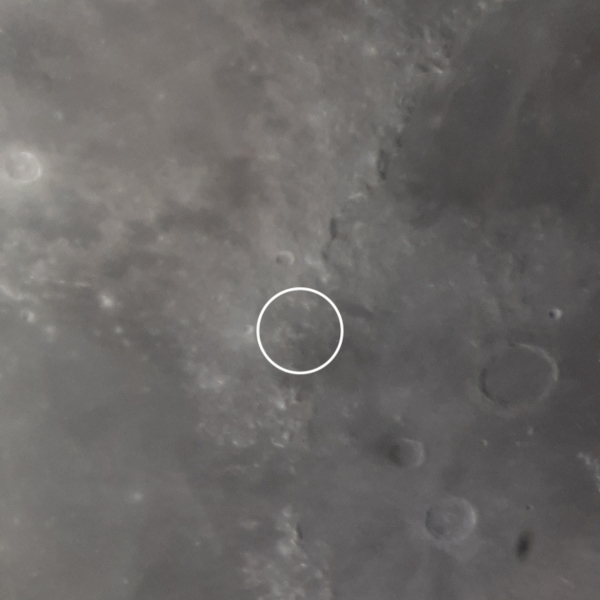}
\hspace{0\textwidth}
\includegraphics[width=0.2\textwidth]{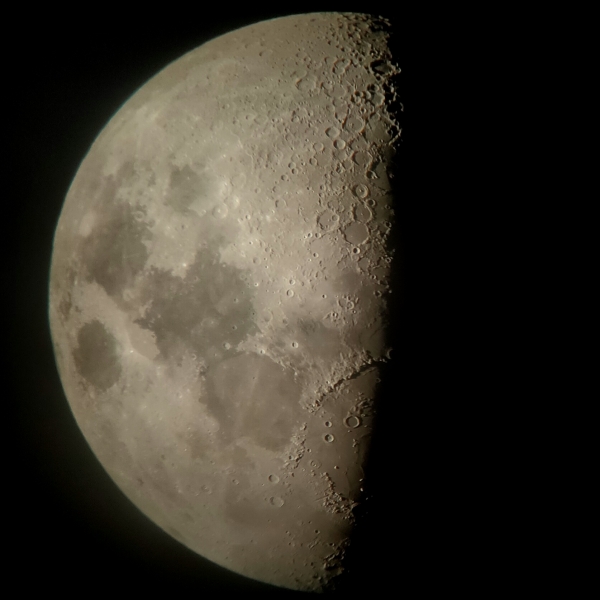}
\hspace{0\textwidth}
\includegraphics[width=0.2\textwidth]{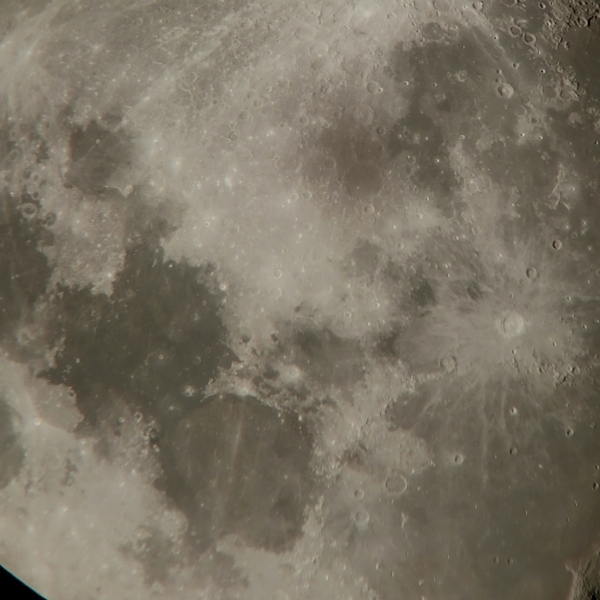}
\hspace{0\textwidth}
\includegraphics[width=0.2\textwidth]{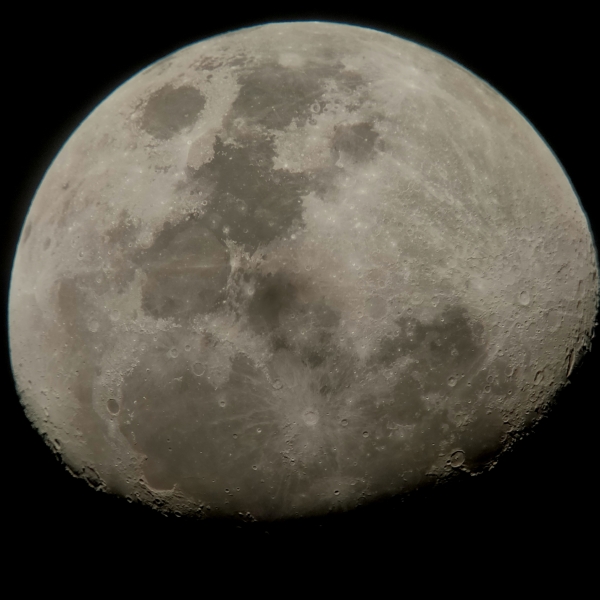}
\hspace{0\textwidth}
\includegraphics[width=0.2\textwidth]{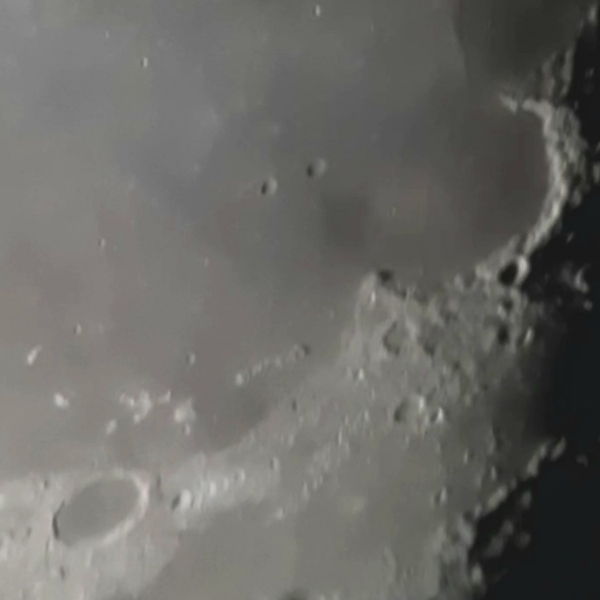}
\caption{Few high resolution pictures of the lunar surface captured by the members. The circles in the images indicate the approximate locations of Apollo landing sites. In the 1st image the circles correspond to 12 and 14 missions, in 2nd image the circle correspond to 15 mission.}
\end{figure}

The observations were conducted at the lush green campus of the Seven Square Academy, Naigaon located on the outskirts of the Mumbai Metropolitan Region \citep{ssan}. The coordinates of the location was 19.362,72.860. All the observations were made during the evenings from 18:00 hrs to 20:30 hrs on different dates in the months of October and November when the Moon was visible above the horizon. 

Student first aligned the 1200mm/150mm Bresser Messier reflector Newtonian telescope's equatorial mount with the polar axis then using a series of high powered eyepieces observed the Moon \citep{bresser/messier}. The eyepiece specifications are as follows: 26 mm Plössl lens providing 46x zoom, 15 mm Plössl lens providing 80x zoom, 9 mm Plössl lens providing 133x zoom. These eyepieces were then combined by a 2x Barlow lens to multiply the power by twice with the maximum power achieved being 266x. 

By conducting the observations the overall observational skills of the members increased as they got acquainted with the various "observable" geological features of the Moon like mare, mountain ranges, craters, lunar rays, mountainous land forms etc. 

The members conducted such visual inspection of the lunar surface with the telescope. \textbf{An example} of \textbf{observational metric }detailed by one of the members \textbf{Priyamvada Shukla} is given here. 

\subsection{Observation Metric}
\subsubsection{Location, Date}
- Seven Square Academy, Naigaon, 29/10/2023.

\subsubsection{Telescope, Eyepieces} 

\begin{enumerate}
    \item What type of telescope are you using?\\
-   We are using a Bresser Messier 150mm diameter, 1200 mm focal length Newtonian reflector telescope \citep{bresser/messier} with 26mm, 15mm, 9mm Plössl eyepieces and 2x Barlow lens.
    
    \item How did you align your telescope for the lunar observations?\\
-   I aligned the telescope on the equatorial mount by aligning it towards the Polaris star direction without the presence of that star. 
\end{enumerate}

\subsubsection{Weather} 

\begin{enumerate}
    \item Describe the current weather condition and presence of clouds.\\
-   The weather was good and it was humid with the presence of a few clouds sporadically.
\end{enumerate}

\subsubsection{Lunar Phase/Condition} 
\begin{enumerate}
    \item What was the current phase of the Moon? Was it bright enough? Describe the shape.\\
-  The phase of the Moon was full Moon and the Moon appeared bright enough. When I observed the Moon with the telescope it was like a bright circle.
\end{enumerate}

\subsubsection{Lunar Features Observed} 
\begin{enumerate}
    \item List the number of prominent craters you observed.\\
-   I observed six prominent craters.

    \item Write the characteristics of the craters in detail.\\
-  Craters appeared to be round shaped, flat on the bottom and appeared very bright because of sun light reflecting off them. Most of the biggest craters glow brightly and couldn't be easily distinguished because of the absence of shadows. Some craters also cast out ray like features in the adjoining regions which appear like light rays emerging from them.

    \item List the number of mare (dark areas) you observed.\\
-  I observed seven mare areas (dark areas).

    \item Write the characteristics of mare areas in detail.\\
-  Lunar mare areas are large and dark regions. They are mostly distributed in a given region of the near side of the Moon. The dark color that appears on the Moon are actually covered with basalt from past volcanic eruptions.

    \item Did you observe any mountain ranges/mountains/elevated regions? How many? \\
-  I observed only one ring like mountain range surrounding a mare region because the shadows casted by the elevated regions were not properly visible.

    \item Write characteristics of the mountain ranges/mountains/elevated regions.\\
-  Mountainous land forms on the Moon appear like elevated areas because of the sparse shadows they cast on the nearby regions. They are often found near to the craters and the boundaries of mare regions. They appeared to decorate one of the mare regions like rings.
    
\end{enumerate}

\subsubsection{Surface Texture and Color} 

\begin{enumerate}
    \item Describe any noticeable texture or color variations on the lunar surface.\\
-  As the Moon was near the horizon it was very red in color but as it moved away from the horizon the color of the surface changed to creamish white.
\end{enumerate}

\subsubsection{Unique or Unusual Features} 
\begin{enumerate}
    \item  Did you observe any unique or unusual features on the Moon's surface?\\
-  The mare regions are more or less connected to each other forming a giant landmass.

    \item  Were there any surface irregularities or formations that caught your attention?\\
-  Because of the full Moon phase the surface didn't appear very irregular because of lower shadows visible.
\end{enumerate}

\subsubsection{Apollo Sites Observations} 

\begin{enumerate}
    \item How many Apollo sites were observed?\\
-   I observed 3 Apollo sites: 12, 14 and 15.

    \item Were they located in craters, mountains, valleys or mare?\\
-   I observed Apollo sites near to the craters and mare regions. The site numbers 12 and 14 were located near Copernicus crater. The site number 15 was located near a mountain range besides Archimedes crater.
\end{enumerate}

\subsubsection{Personal Experiences} 
\begin{enumerate}
    \item How did you feel during the observation? What was your overall impression of the Moon?\\
-  I was very excited while observing the Moon as I observed the celestial body for the first time. I learnt about craters, mare regions, mountains and valleys on the Moon and understood them better by viewing them live through a telescope.

    \item What did you find most interesting or captivating about your lunar observation?\\
-  I found most interesting about the surface of the Moon is that it is filled with scenic features which otherwise is not possible to be viewed without a telescope.
\end{enumerate}

\subsubsection{Other Planets} 
\begin{enumerate}
    \item Did you observe any planets?\\
- Yes, I observed Saturn and Jupiter. I also observed five moons of Jupiter and one moon of Saturn.

    \item Write in brief about their appearance i:e appearance, presence of moons, surface texture.\\
- The color of Jupiter was yellowish. I saw the outermost layer of its clouds but the great red spot wasn't visible. The five moons of Jupiter appeared like stars and same was with one moon of Saturn. I observed Saturn as well. The color of Saturn was creamish. I observed its rings. The rings of Saturn appeared clean and shiny.
\end{enumerate}

\section{Data Analysis}

\begin{figure}[ht!]
\centering
\rotatebox{0}{\includegraphics[width=0.4\textwidth]{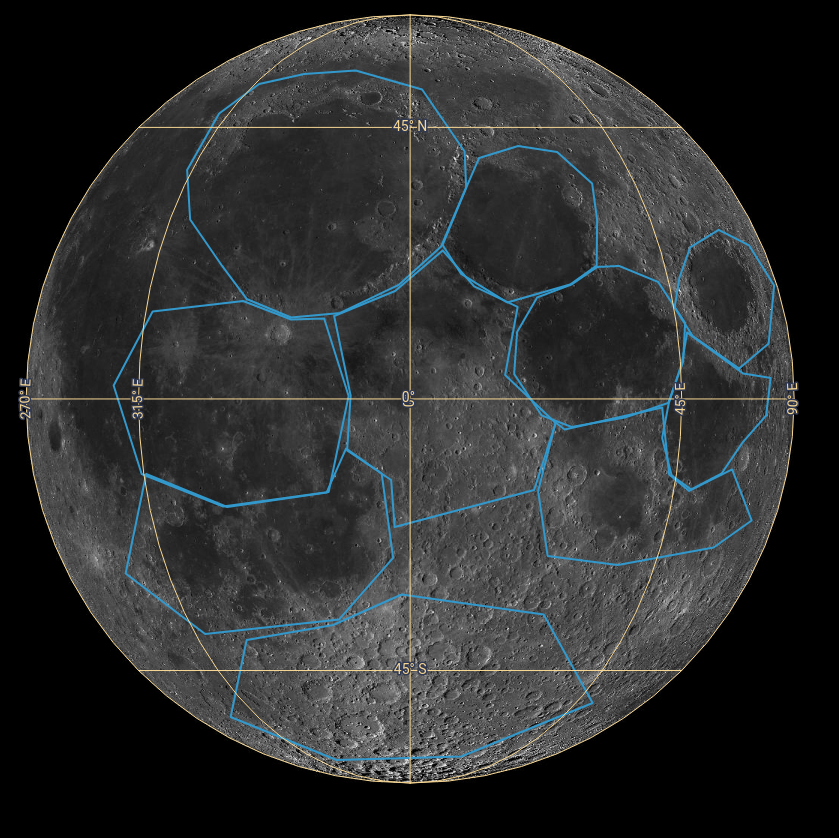}}
\caption{The ten lunar zones for data analysis assigned to the members from the LROC quickmap. Credit: LROC quickmap \citep{lroc_quickmap}.}
\end{figure}

After the observations were complete members were assigned different lunar zone from the LROC quickmap for visual analysis of topographical features.

In the following subsections \textbf{one example} of the \textbf{data analysis metric} for one of the zones filled by member \textbf{Abhinav Mishra} is given.

\subsection{Data Analysis Metric}

\subsubsection{Topographical Analysis from LROC Map}
The topographical analysis from the WAC and NACs from LROC \citep{lroc_quickmap} is given here.

\begin{enumerate}
    \item Write the bounding coordinates of the zone and provide the surface area of the zone.\\
- The bounding coordinates of this zone [Lat, Lon] are: [26.14281, 63.57576], [23.23148, 52.44202], [19.14796, 48.32422], [18.54230, 47.78237], [13.80300, 45.17245], [9.41772, 48.24169], [4.57381, 59.28972], [8.29515, 70.70602], [17.19758, 83.09212], [23.57237, 74.40981]. The surface area of the zone is 452724.970 km².

    \item How does the zone appear to be?\\
- The zone has  Mare Crisium in the center having mostly smooth surface with some craters of diameter of around 20 kms inside. The mare is bounded by a distinct ring of elevated regions and hills and the rest of the area around the elevated regions is filled with some craters. The Mare Crisium is also lined by some ridges called dorsa also. There is another adjacent mare region called as Mare Anguis located near Eimmart craters. Like the bigger mare, the smaller Mare Anguis is surrounded by elevated regions. The zone appears grayish black. The zone is not too bright nor too dark.

    \item  Is it a mare region? If yes then write the name of the mare region, describe the mare region, provide coordinates and the surface area.\\
- Yes the given zone is a mare region. The name of the mare is Mare Crisium. It is mostly circular in shape and the surface inside appears to be mostly smooth in WAC map but with NACs map several tiny craters inside like Peirce, Picard, Lick, Yerkes and a few ridges in the inner part called dorsa like Harker, Tetyaev, Oppel, Termier etc are visible. The bounding coordinates [Lat, Lon] of the bigger mare region is as follows: [24.07145, 59.73956], [23.07262, 56.39623], [21.20880, 52.51672], [17.41111, 49.88751], [14.09506, 49.83230], [9.97559, 56.18770], [10.90648, 62.24676], [15.39333, 69.14022], [19.75685, 69.14496], [23.95691, 62.30749]. The surface area of the Mare Crisium is 172494.476 km². 

    \item Are there any prominent  mountains, mountain ranges in the zone? Provide coordinates.\\
- There is only one named mountain in the given zone called Mons Usov at coordinates [Lat, Lon] of [11.92908, 63.26895]. But there are a lot of elevated mountainous land forms and  hills in the zones at the southern part of the zones and most of the elevated regions line the central mare region. The coordinates [Lat, Lon] of the mountainous land forms lining the central mare region are as follows: [21.21890, 51.36163], [19.06761, 49.75241], [17.75606, 49.04479], [15.56269, 48.13944], [14.29591, 48.01610], [12.73499, 48.92166], [11.75387, 50.80665], [9.64768, 56.02920], [9.91679, 59.93023], [10.31051, 62.46678], [10.85008, 63.48655], [14.06784, 65.97195], [12.99772, 66.90363], [14.93175, 70.32258], [16.45746, 70.64409], [18.16695, 70.42033], [19.76324, 70.09685], [20.61091, 67.93510], [22.01237, 65.81380], [24.20829, 64.30417], [25.19631, 62.65011], [24.67887, 58.64200], [23.47229, 55.19431], [22.43106, 52.81553]. 

    \item How many prominent craters  are there in the zone?\\
- There are around 16 prominent craters in the zone. Some of their names and coordinates are as follows: \\
a. Picard:		[14.60401, 54.72482]\\
b. Condorcet:	[12.11209, 69.68968]\\
c. Alhazen:	    [15.92418, 71.90608]\\
d. Hansen:	    [14.0520, 72.26641]\\
e. Auzout:		[10.23645, 64.04636]\\ 
f. Shapely:	    [9.42083, 56.80144]\\ 
g. Lick:		[12.39289, 52.78271]\\
h. Firmicus:    [7.26789, 63.33301]\\

    \item If you see any crater that is prominent or very big then write about the appearance/topography of the crater.\\
- Crater Condorcet: The Crater Condorcet at central coordinates [Lat, Lon] [12.11209, 69.68968] is a large crater having the surface area of 4363.966 km² and is surrounded by many tiny adjoining craters to it like Condorcet T, A, H etc. It appears to be oval in shape and is located at the southern part of the given region. The inner region of the crater appears smooth from far off but is covered with a large number of very small craters. There is also a small unnamed crater inside which is visible at the right hand side located at coordinates [Lat, Lon]: [12.11776, 70.30992] with a radius of approximately 1.8 kms.

    \item  Is there any Apollo site located in the zone? If yes then write the coordinates of the sites.\\
- There is no Apollo site present in this region.

    \item Describe the Apollo site very shortly (If the Apollo site is not present, skip this question).\\
- Not available.

    \item Are there any man made objects/spacecrafts/landers in the region? If yes then provide the names and their coordinates.\\
- There are 2 spacecrafts located in the region.\\
  a. Luna 23 -	[12.66674,62.15112]\\
  b. Luna 24 -	[12.711424,62.21295]
  
    \item If there are man made objects then describe very shortly the region. (If there is no man made object then skip this question).\\
- The two  spacecrafts are located really very close to each other and their landing sites are relatively flat but filled with very small craters. Both of them are present inside the Mare Crisium.

\end{enumerate}

\begin{figure}[ht!]
\centering
\rotatebox{0}{\includegraphics[width=0.3\textwidth]{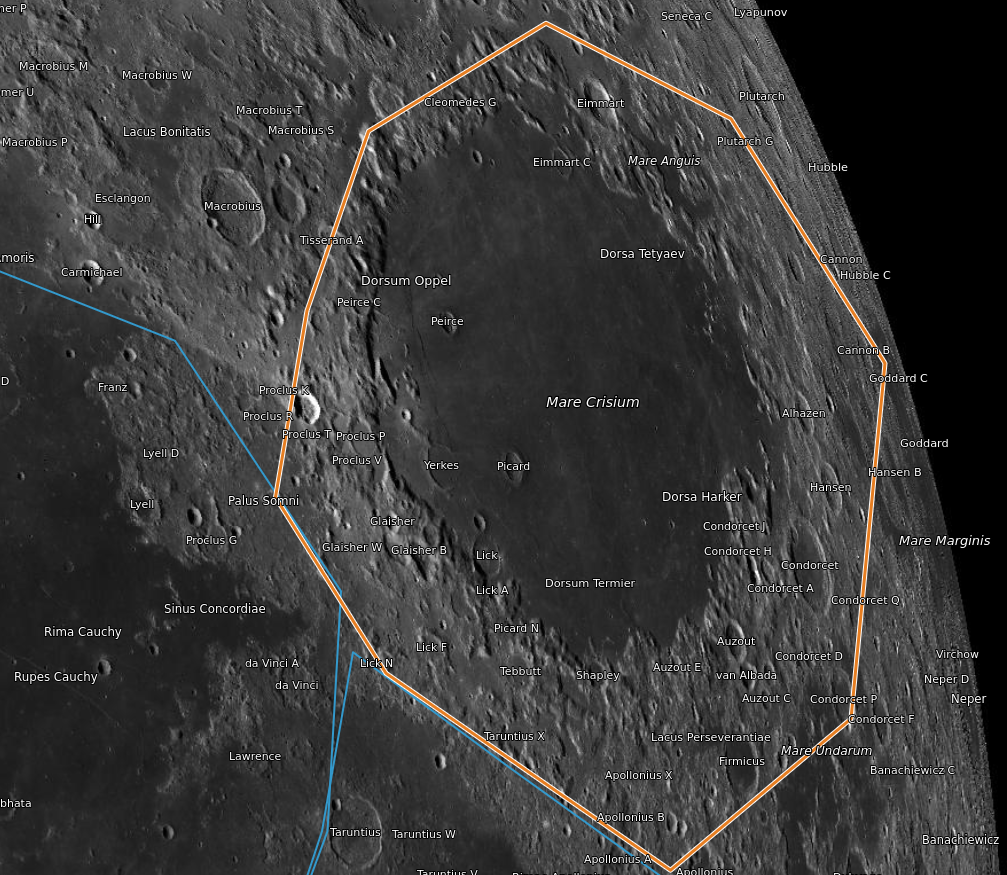}}
\hspace{0\textwidth}
\rotatebox{0}{\includegraphics[width=0.3\textwidth]{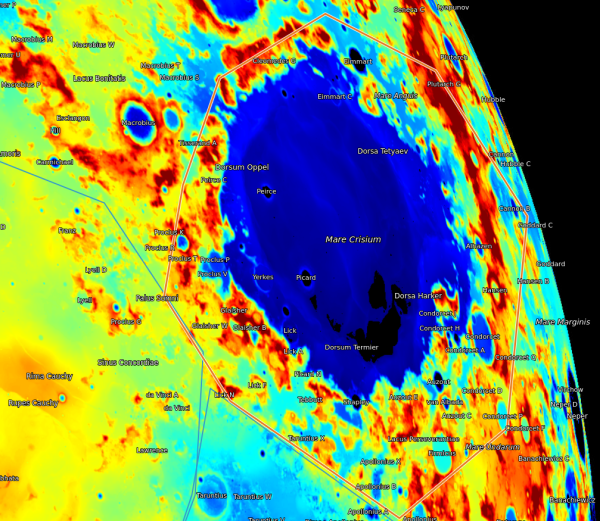}}
\hspace{0\textwidth}
\rotatebox{0}{\includegraphics[width=0.3\textwidth]{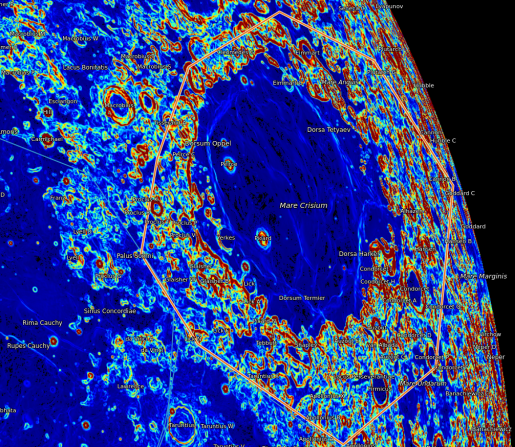}}
\hspace{0\textwidth}
\rotatebox{0}{\includegraphics[width=0.3\textwidth]{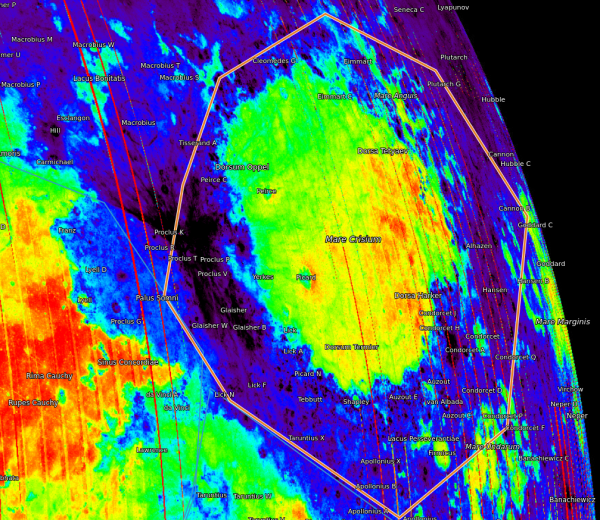}}
\hspace{0\textwidth}
\rotatebox{0}{\includegraphics[width=0.3\textwidth]{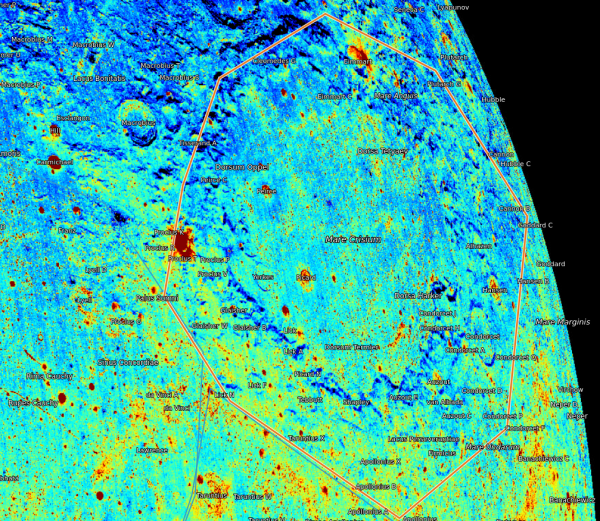}}
\caption{Different layers of the lunar zones for analysis from one of the 10 zones. From clockwise 1st image: WAC image of the zone, Credit: LROC quickmap \citep{lroc_quickmap}, 2nd image: Height variation data, Credit: LROC quickmap elevation data \citep{barker2015lunar, henriksen2020lroc, lroc_quickmap}, 3rd image: Slope variation data, Credit: LROC quickmap slope data \citep{barker2015lunar, henriksen2020lroc, lroc_quickmap}, 4th image: Regolith temperature variation data, Credit: LROC quickmap Diviner data, \citep{paige2010lro, lroc_quickmap, powell2023high}, 5th image: Iron oxide abundance data from Kaguya, Credit: LROC quickmap Kaguya data, \citep{kato2009kaguya, lemelin2019compositions, lroc_quickmap}.}
\end{figure}

\subsubsection{Terrain Height Variations}
The terrain height variation data on LROC quickmap \citep{lroc_quickmap} is collected from a series of combined data products from LOLA of LRO and Terrain Camera of Kaguya \citep{barker2015lunar, henriksen2020lroc}.

\begin{enumerate}
    \item What is the overall variation of color and height of the layer on the zone? \\
- Overall the color of the zone appears to be blue in the central region and brownish red, yellowish orange in the outer regions. Inside  the mare region  the color appears  mostly  blue indicating lower elevation than other areas, some black region are also observed which indicates the lowest elevation in the given the given region. We can observe that the mare is bounded by brownish red and yellowish orange areas indicating that the mare is covered with high elevation area, usually a mountain, a mountain range, a mountainous land form , hills and rims of craters. The dorsa features in the Mare Crisium have very low elevations as compared to outer mountainous land forms.

    \item What do the regions with shades of red correspond to? Explain in detail?\\
- The regions with shades of red correspond to elevated surfaces like mountainous land forms, hills, mountain ranges or rims of craters. The belt of mountainous land forms near the Mare Anguis extending to Condorcet crater is especially dark brownish in color. This indicates that the elevation is highest in that belt of region. There are similar such highest elevation regions on the other side of the mare region at Glaisher craters series, Cleomedes G crater, Tisserand A crater and Proculus crater series but they are disconnected pockets of mountainous land forms. The average elevation in those regions is around 1600 m and highest elevation is around 2200 m.

    \item  What do the regions with shades of blue correspond to? Explain in detail? \\
- The regions with shades of blue correspond to regions with lower elevation in the Mare  Crisium and the floors of the craters. The Mare Anguis has lighter shade of blue as compared to Mare Crisium indicating higher elevation compared to Mare Crisium. The floors of the craters like Peirce and Picard inside the Mare Crisium have black color indicating lowest elevation, while the floors of the craters out of the mare Crisium have lighter shade of blue as compared to the mare indicating the floors of the craters outside the mare region have higher elevation than the elevation within the mare region. A black patch is clearly visible in the lower end of the Crisium  near Dorsa Termier and Dorsa Harker indicating the region/point of lowest elevation in the entire zone. The average elevation in those regions is around -3500 m and lowest elevation is around -3888 m. 
\end{enumerate}

\subsubsection{Terrain Slope Variations }
The terrain slope variation data from LROC quickmap \citep{lroc_quickmap} is obtained from terrain height analysis from a series of combined data products from LOLA of LRO and Terrain Camera of Kaguya \citep{barker2015lunar, henriksen2020lroc}.

\begin{enumerate}
    \item What is the overall variation of color of the layer on the zone?\\
- In the inner region the colors are blue and shades of blue and as the distance from the inner region increases the color changes to yellowish, green shades to dark brownish, red shades. The overall color in the area around the boundaries of Mare Crisium is brownish red indicating that the surface there has steeper slope in comparison to other areas of the region. The rims of the craters have comparably steep slopes along with the slopes of mountainous land forms which have higher slope values. While the surface inside the mare region appears to be bluish with some blackish spots inside indicating that the surface is comparatively smoother with very low slope. 

\end{enumerate}

\subsubsection{Iron Oxide Abundance}
The iron oxide data is obtained from  Kaguya mission's SP analysis \citep{lemelin2019compositions}.

\begin{enumerate}
    \item What is the overall variation of color of the layer on the zone?\\
- Overall the color variation  follows a trend from outside to inside wherein the color at the outer regions of the zone appears to be black, purple and darkish blue while the area inside the mare region appears green to orange. This shows that the outer parts of the zone are poor in iron oxide while the central part of the zone is rich in iron oxide.

    \item How much is the approximate abundance in the regions where the abundance appears to be high?\\
- The abundance of iron oxide in those regions is around 19 percent wt. 

    \item How much is the approximate abundance in the regions where the abundance appears to be low?\\
-  The abundance of the iron oxide in those regions is around 5 percent wt.

\end{enumerate}

\subsubsection{Regolith Temperature}
The regolith temperature map has been constructed from the Diviner data \citep{powell2023high}.

\begin{enumerate}

    \item What is the most dominant and least dominant color observed? \\
- Sky blue and shades of blue are the most dominant colors in the given region with very small pockets of dark brown to red color present in some craters. 

    \item Is there any variation in the color?\\
    - There is no variation in color observed.
    
    \item What is the average value of regolith temperature?\\
    - The average value of regolith temperature is around 100 K.
\end{enumerate}

\section{Discussion}
The project done by us is the major project undertaken by Science-AI Symbiotic Group \citep{science_ai_symbiotic_group} in the field of astronomy. The detailed lunar observation metric text and the detailed data analysis metric text compiled by us will be released soon in future.

The lunar surface observation project has demonstrated a significant positive impact on the participating students' education. By actively engaging in real-time observations and subsequent data analysis, the students have gained a deeper understanding of astronomy and geology. The hands-on experience with the Bresser Messier reflector Newtonian telescope has ignited a lasting interest in observational science. It would be beneficial to assess any improvements in the students' academic performance and their sustained enthusiasm for science-related subjects in the long term.

One of the remarkable aspects of this project is its cost-effectiveness. The utilization of the Bresser Messier reflector Newtonian telescope \citep{bresser/messier}, a relatively affordable tool, makes this model accessible to schools with limited resources. The reliance on open-access data from the LRO ensures that the project maintains a low budget while providing authentic and reliable information. This cost-effective approach not only enhances the project's replicability but also serves as a beacon for schools seeking impactful yet economical science education initiatives.

The project has the potential to serve as a compelling science outreach program. Organizations interested in popularizing planetary sciences research at the school level could leverage this model. The appeal of studying the lunar surface in real-time captures the imagination of students and the broader community. Collaborations with external organizations or scientists could further enhance the outreach impact, fostering a sense of community engagement with scientific endeavors. This outreach aspect could be a catalyst for generating interest in STEM fields beyond the participating students.

The hands-on nature of the project has undoubtedly contributed to the development of crucial skills among the students. From honing their data analysis skills using LROC's open-access data to fostering teamwork and critical thinking during the topographical analysis, the students have gained practical experience aligned with real-world scientific practices. These skills extend beyond the realm of planetary sciences and are transferable to various academic and professional pursuits. Future iterations of the project could consider incorporating more intentional skill development components to further enhance the educational impact.

Various important concepts of astronomy like recognising various celestial objects, recognising the significance of movement of earth about its axis, recognising geological features on other planetary objects and the sky changes over seasons cannot be properly understood by just watching videos, studying from books etc. But on the other hand using a telescope, organising the alignment of the telescope in the field and then the help of various substitute instruments/technologies like compass, gps location can significantly boost the acquisition of the knowledge of the sky and the astronomy in general. Hence it was a project to experiment with students/members to test the efficiency of field work and computer work. 

Students spend considerable time in the context of Indian education system to grasp various concepts of geography and earth science. The visual observation of the maps available in textbooks doesn't impact the learning process in the students effectively. Students respond particularly well to learning when interactive simulations, maps and visually appealing images, plots and graphs are shown. Hence learning with the help of open source data and field work shows to boost the learning potential of the students at secondary level. 

Looking ahead, there are exciting possibilities for the Science-AI Symbiotic Group \citep{science_ai_symbiotic_group}. Collaborations with other schools, both locally and globally, could broaden the scope and impact of future projects. Furthermore, integrating emerging technologies or methodologies could provide students with exposure to cutting-edge advancements in space exploration. Continuous innovation and adaptability will be key in ensuring the sustained success and relevance of the group's initiatives.

\section{Software}
We have made extensive use of data products from LRO and Kaguya missions which are collectively hosted on the LROC quickmap website \citep{lroc_quickmap}. 

\section{Conclusion}
Our project involving the studies of lunar surface with telescope and open access data is fruitfully completed. The necessary supplementary material will be released in future. The project has laid groundwork for more interesting short duration projects in the context of astronomy and data analysis. The Science-AI Symbiotic Group will be working towards more interesting projects where science and AI will intersect. 

\section{Acknowledgment}
Science-AI Symbiotic Group is grateful to Principal Mrs. Deepinder  Chadha of the Seven Square Academy, Naigaon for the access to the school infrastructure and for being the catalyst behind the establishment of this research group. We are grateful to the Seven Eleven Education Society and its Chairperson Mrs. Suman Narendra Mehta for providing the school valuable inspiration to propel towards knowledge and impactful education. We are grateful to the parents for cooperating during the duration of the project.

\bibliography{moon_paper}

\begin{thebibliography}{}
\expandafter\ifx\csname natexlab\endcsname\relax\def\natexlab#1{#1}\fi
\providecommand{\url}[1]{\href{#1}{#1}}
\providecommand{\dodoi}[1]{doi:~\href{http://doi.org/#1}{\nolinkurl{#1}}}
\providecommand{\doeprint}[1]{\href{http://ascl.net/#1}{\nolinkurl{http://ascl.net/#1}}}
\providecommand{\doarXiv}[1]{\href{https://arxiv.org/abs/#1}{\nolinkurl{https://arxiv.org/abs/#1}}}

\bibitem[{sci(2023)}]{science_ai_symbiotic_group}
 2023, Science-AI Symbiotic Group.
\newblock \url{https://science-ai-symbiotic-group.github.io/}

\bibitem[{ssa(n.d.)}]{ssan}
 n.d., Seven Square Academy, Naigaon.
\newblock \url{https://naigaon.sevensquareacademy.org/}

\bibitem[{bre(n.d.)}]{bresser/messier}
 n.d., Bresser Messier.
\newblock \url{https://www.bresser.de/en/Astronomy/Telescopes/BRESSER-Messier-NT-150L-1200-Hexafoc-EXOS-2-EQ5-Telescope.html}

\bibitem[{Barker {et~al.}(2015)Barker, Mazarico, Neumann, Zuber, Haruyama, \& Smith}]{barker2015lunar}
Barker, M.~K., Mazarico, E., Neumann, G.~A., {et~al.} 2015, Icarus, 273, 346

\bibitem[{Henriksen {et~al.}(2020)Henriksen, Manheim, Robinson, \& {LROC Team}}]{henriksen2020lroc}
Henriksen, M.~R., Manheim, M.~R., Robinson, M.~S., \& {LROC Team}. 2020, in Lunar Surface Science Workshop 2020 (LPI Contrib. No. 2241), Lunar Reconnaissance Orbiter Camera Science Operations Center, Arizona State University, Tempe, AZ 85287

\bibitem[{JAXA(n.d.)}]{jaxa_kaguya}
JAXA. n.d., JAXA/Kaguya.
\newblock \url{https://www.isas.jaxa.jp/en/missions/spacecraft/past/kaguya.html}

\bibitem[{Kato {et~al.}(2009)}]{kato2009kaguya}
Kato, M., {et~al.} 2009, Acta Astronautica, 65, 1481

\bibitem[{Lemelin {et~al.}(2019)Lemelin, Lucey, Miljković, Gaddis, Hare, \& Ohtake}]{lemelin2019compositions}
Lemelin, M., Lucey, P.~G., Miljković, K., {et~al.} 2019, Planetary and Space Science, 165, 230

\bibitem[{LRO(n.d.)}]{lro_nasa}
LRO. n.d., Lunar Reconnaissance Orbiter (LRO) - NASA.
\newblock \url{https://lunar.gsfc.nasa.gov/about.html}

\bibitem[{LROC(n.d.)}]{lroc_website_closeup}
LROC. n.d., LROC Website - LROC Camera Closeup.
\newblock \url{http://lroc.sese.asu.edu/posts/139}

\bibitem[{MSS(n.d.)}]{msss_lro_camera}
MSS. n.d., Malin Space Science Systems - LRO Camera.
\newblock \url{https://www.msss.com/all_projects/lro-camera.php}

\bibitem[{Paige {et~al.}(2010)}]{paige2010lro}
Paige, D.~A., {et~al.} 2010, Space Science Reviews, 150, 125

\bibitem[{Powell {et~al.}(2023)Powell, Horvath, Robles, Williams, Hayne, Gallinger, \& et~al.}]{powell2023high}
Powell, T.~M., Horvath, T., Robles, V.~L., {et~al.} 2023, Journal of Geophysical Research: Planets, 128, e2022JE007532

\bibitem[{Quickmap(n.d.)}]{lroc_quickmap}
Quickmap. n.d., LROC Quickmap.
\newblock \url{https://quickmap.lroc.asu.edu}

\bibitem[{Robinson {et~al.}(2010)}]{robinson2010lro}
Robinson, M.~S., {et~al.} 2010, Space Science Reviews, 150, 81

\end{thebibliography}
\bibliographystyle{aasjournal}

\appendix
\section{Apollo Landing Sites: A Brief Observational Overview}
Neil Armstrong (Apollo 11), Buzz Aldrin (Apollo 11), Charles "Pete" Conrad (Apollo 12), Alan Bean (Apollo 12), Alan Shepard (Apollo 14), Edgar Mitchell (Apollo 14), David Scott (Apollo 15), James Irwin (Apollo 15), John W. Young (Apollo 16), Charles Duke (Apollo 16), Eugene Cernan (Apollo 17), Harrison Schmitt (Apollo 17) have landed on the Moon. Apollo landing sites have been historically one of the greatest events of controversy among the general public because of the ignorance of understanding the truth supplemented by very little data available in the late 70s, 80s, 90s from various sources. In an interview conducted by the members of Science-AI Symbiotic Group it was found that there is a sizeable amount of students by around 10 percent at school level from grades 1 to 10 that still believe that the landings were an hoax. It can thus be guessed whether how much disinformation and propaganda is still out there among general population about the single most greatest achievement in the history of humans. 

There has been ample evidence from various space probes to discredit the fake news which is still evident in the society about the Moon landings. In the series of high resolutions images the evidence of Moon landings as confirmed by LROC's high resolution images is presented \citep{lroc_quickmap}. 

\subsection{Apollo Sites as seen from NACs }

\begin{figure}[ht!]
\centering
\rotatebox{0}{\includegraphics[width=0.3\textwidth]{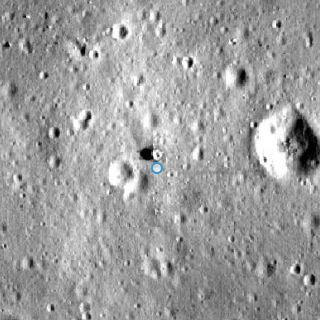}}
\hspace{0\textwidth}
\rotatebox{0}{\includegraphics[width=0.3\textwidth]{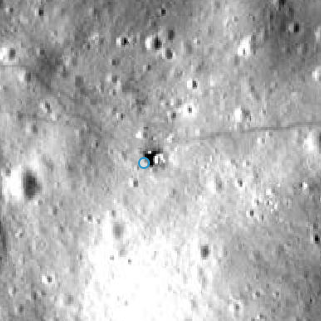}}
\hspace{0\textwidth}
\rotatebox{0}{\includegraphics[width=0.3\textwidth]{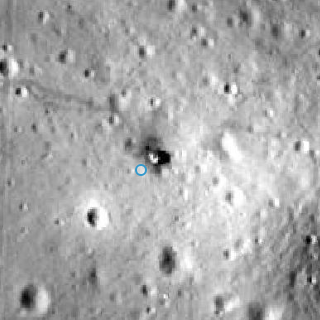}}
\hspace{0\textwidth}
\rotatebox{0}{\includegraphics[width=0.3\textwidth]{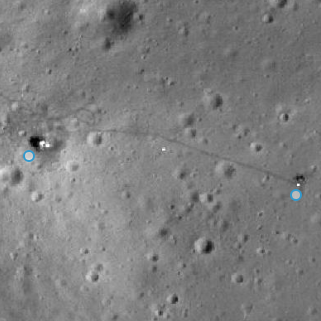}}
\hspace{0\textwidth}
\rotatebox{0}{\includegraphics[width=0.3\textwidth]{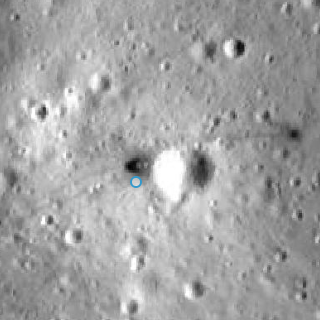}}
\hspace{0\textwidth}
\rotatebox{0}{\includegraphics[width=0.3\textwidth]{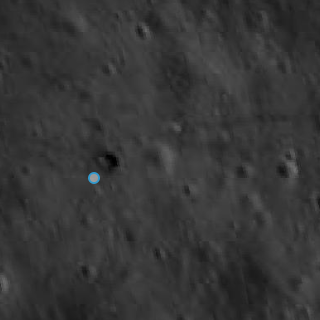}}

\caption{In clockwise 1st image: Apollo 11, 2nd image: Apollo 12, 3rd image: Apollo 14, 4th image: Apollo 15, 5th image: Apollo 16, 6th image: Apollo 17 landing sites. Credit: NACs from LROC quickmap \citep{lroc_quickmap}.}
\end{figure}

In the NACs images, distinct objects/anthropogenic features at the site of Apollo 11, 12, 14, 15, 16 and 17 missions are visible. Blue colored micro points have been dropped at the zone where the landing modules (LMs) are present. Additionally the lunar roving vehicles (LRVs) are also visible in the later missions of 15, 16, 17 next to the LMs.

\end{document}